\pdfoutput=1
\documentclass[letterpaper, 10 pt, conference]{ieeeconf}  

\IEEEoverridecommandlockouts                              
\usepackage{graphicx} 
\usepackage{subfigure}
\usepackage{amsmath} 
\usepackage{amssymb}  
\usepackage{autobreak}
\usepackage{xcolor}
\definecolor{mygreen}{rgb}{0.466, 0.674, 0.188}

\let\labelindent\relax 
\usepackage{enumitem}

\usepackage{siunitx}

\usepackage[ruled,linesnumbered]{algorithm2e}

\title{\LARGE \bf
Autonomous Parking of Vehicle Fleet in Tight Environments
}

\author{Xu Shen, Xiaojing Zhang, Francesco Borrelli
\thanks{X. Shen, X. Zhang and F. Borrelli are with the Model Predictive Control Lab, Department of Mechanical Engineering, 
        University of California, Berkeley, USA.
        Email: {\tt\small \{xu\_shen, xiaojing.zhang, fborrelli\} at berkeley.edu}}%
}

\begin{document}

\maketitle
\thispagestyle{empty}
\pagestyle{empty}

\begin{abstract}


The problem of autonomous parking of vehicle fleets is addressed in this paper. 
We present a system-level modeling and control framework which allows investigating different vehicle parking strategies while taking into account path planning and collision avoidance.
The proposed approach decouples the problem into a centralized parking spot allocation and path generation, and a decentralized collision avoidance control. 
This paper presents the hierarchical framework and algorithmic details. Extensive simulations are used to assess several allocation strategies in terms of total fleet parking time and queue length. In particular, we observe that when parking large vehicle fleets,
a phenomenon similar to Braess's paradox occurs.
\end{abstract}

\section{Introduction}
Vehicle parking is becoming increasingly more challenging for drivers. According to the statistics in~\cite{McCoy2017}, drivers in New York City spend an average of 107 hours a year searching for parking spots. Increased population density inevitably reduces parking availability. As a result, drivers are having to deal with long queues and tighter spaces while entering parking facilities~\cite{Banzhaf2017_2}. This increases not only the complexity of the parking maneuver, but also the choice of the parking spot along the way~\cite{Bertsekas2018}.

The application of automated  vehicles (AVs) technology is becoming ubiquitous, including automated valet parking system~\cite{Li2019}. However, when a large fleet of AVs is trying to park, the interaction among vehicles will become more complex and parking allocation strategies need to be studied to guarantee system efficiency. In addition, Vehicle-to-Vehicle (V2V) and Vehicle-to-Infrastructure (V2I) communication~\cite{Guanetti2018,Turri2017,Liu2018} has the potential to improve system efficiency.

Researches have investigated the occupancy and waiting time estimation problem in parking scenarios of AVs in the past. In the majority of existing literature, the vehicles are modeled as a service queue~\cite{Arnott1999,Tavafoghi2019} or traffic flow~\cite{Pidd1996} where individual kinematic constraints and the body geometry are ignored. Although these simplifications facilitate the study of macro-level traffic behavior, they suffer from two major limitations: (\romannumeral1) The parking trajectories may be dynamically infeasible for an actual vehicle; (\romannumeral2) The inter-vehicle interaction is neglected so that it cannot be adapted to scenarios where vehicles have maneuvering constraints. These limitations become significant when vehicles need to follow complex trajectories to maneuver into narrow parking spots. However, directly applying current path planning and collision avoidance methods~\cite{Li2019, Wang2014} will be computationally infeasible with the time-varying nonconvex environment configuration and vehicle interactions among the large fleet AVs.

This paper addresses the limitations discussed above for the parking of AVs fleet. In particular, the contributions are:
\begin{enumerate}[label=(\roman*)]
    \item A generic framework is proposed where the parking spot allocation, path planning, and vehicle interaction control are decoupled. A centralized coordinator is only responsible for spot allocation decisions and path generation, while collision avoidance is performed by each vehicle in parallel. 
    \item A numerical efficient implementation of the proposed approach is presented, which avoids the real-time path planning and simplifies the computationally intense safety evaluation among vehicles.
    \item The proposed algorithm is used in extensive simulation to assess several allocation strategies in order to evaluate the total fleet parking time and queue length.
\end{enumerate}

The paper is organized as follows. Section~\ref{sec:setup} introduces the problem scenarios and assumptions; Section~\ref{sec:framework} establishes the architecture design for a generic, computational tractable solution for this type of problem; Section~\ref{sec:proposed_algorithm} presents a control algorithm implementation with allocation strategies, offline path generation and collision avoidance on occupancy grids; Section~\ref{sec:simulation} shows the influence of allocation strategies on the system efficiency using simulation result; Finally, Section~\ref{sec:conclude} concludes the paper.

\section{Problem Setup}
\label{sec:setup}
\subsection{Scenario Description}
\vspace{-0.3cm}
\begin{figure}[ht]
\centering
\includegraphics[width=\columnwidth]{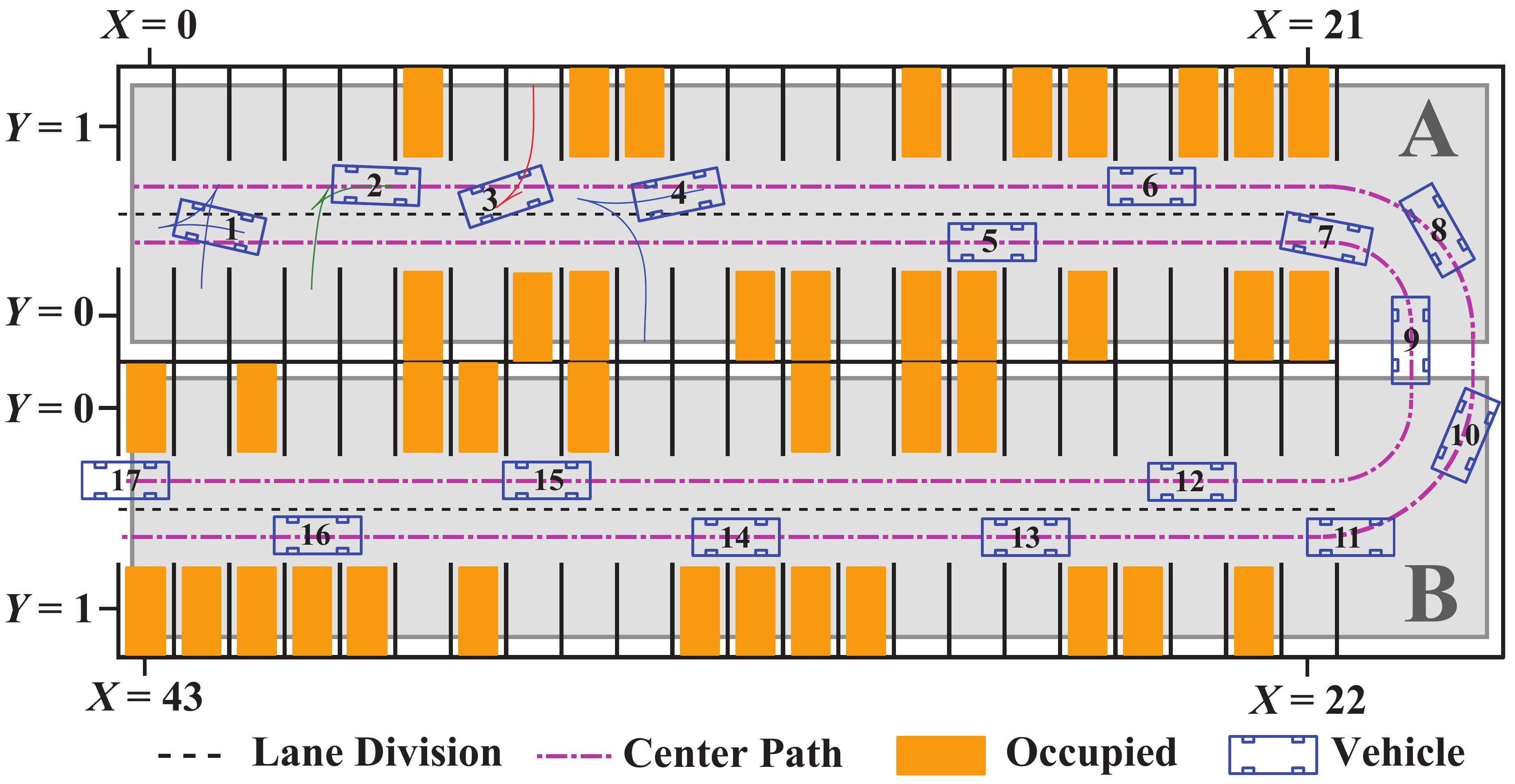}
\caption{Scenario Overview}
\label{fig:Overview}
\vspace{-0.3cm}
\end{figure}

The problem scenario is depicted in Fig.~\ref{fig:Overview}, where a parking lot has an entrance gate at the lower-left corner, an outlet gate at upper left corner, two driving lanes available, and two blocks (A and B) offering 88 parking spots in total. They are partially occupied and a fleet of vehicles is entering from the entrance using two lanes. In this paper, we mainly investigate the influence of incoming fleet and each vehicle's ultimate goal is to find a target parking spot and move into it without colliding. A local $(X,Y)$ frame along the driving lanes is used to uniquely define the position of any parking spot. The spots farthest from the gate are associated with $X=0$, while those next to the inner lane are associated with $Y=0$.

Referring to the above scenario, each autonomous vehicle (AV) undergoes three modes during a complete parking task:
\begin{itemize}
\item \textbf{Allocation} mode: Upon entering, the vehicle's driving lane and the final target parking spot are determined, as shown by vehicle $17$ in Fig.~\ref{fig:Overview}. In human driving, this decision is usually made by vehicle owner or a coordinator of the parking lot; 
\item \textbf{Queuing} mode: The vehicle drives along the selected lane and gradually moves to the target spot while avoiding possible collisions on the way, as shown by vehicles $5-16$ in Fig.~\ref{fig:Overview};
\item \textbf{Maneuvering} mode: The vehicle performs a parking maneuver safely to end up with the specific position and heading inside the designated spot, as shown by vehicles $1-4$ in Fig.~\ref{fig:Overview}.
\end{itemize}

\subsection{Assumptions}

The following assumptions are used throughout this paper:

(A1) Each vehicle is fully-autonomous and instrumented with a communication device to exchange information with the central coordinator. The network bandwidth is sufficient; 

(A2) The time required to compute the spot allocation and vehicle paths is negligible;

(A3) There exists a low-level controller to track the computed path perfectly with a specific speed profile. 

\section{Hierarchical Framework}
\label{sec:framework}

\begin{figure}
    \centering
    \includegraphics[width=\columnwidth]{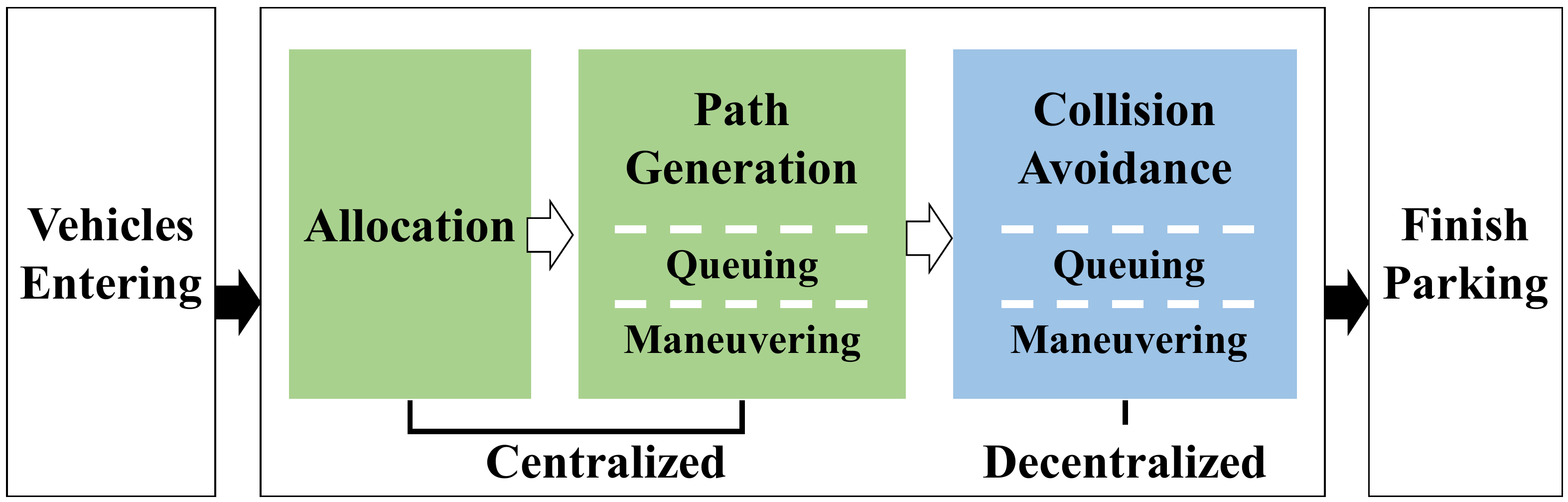}
    \caption{System Framework}
    \label{fig:framework}
    \vspace{-0.4cm}
\end{figure}

We decompose the parking problem into the hierarchical control architecture shown in Fig.~\ref{fig:framework}:

\begin{enumerate}[label=(\roman*)]
\item A central coordinator is in charge of allocating parking spots to vehicles. Clearly the allocation strategy will change the interaction among vehicles and the overall traffic pattern.
\item A centralized path generating algorithm provides trajectories for both queuing and maneuvering modes of vehicle operation. While generating paths, inter-vehicle collision avoidance is not addressed, which reduces the complexity of planning and generalizes the results for repetitive scenarios.
\item Vehicles avoid collision locally in a decentralized manner until they reach the final goal. The safety constraints are defined respectively for vehicles in queuing and maneuvering mode.
\end{enumerate}

\section{Proposed Algorithm}
\label{sec:proposed_algorithm}
In this section we present an algorithm which follows the architecture in Section~\ref{sec:framework}.
The central coordinator allocates vehicles to parking spots and generates paths from an offline library. As for collision avoidance, vehicles make decisions by referring to a shared occupancy grid map, which is distributed to each vehicle and modified synchronously for the decentralized control.

\subsection{Allocation Strategies Design}
\label{sec:allocate_strategy}
We assume that lane opening and spot assignment is controlled at this level. Since two lanes are available in the scenario we study (Fig.~\ref{fig:Overview}), the algorithm can decide to open just one lane (1L) or both two lanes (2L) for incoming vehicles. Regardless of the lane opening strategy, we design three spot assignment policies. The first one is a random baseline. The second and third search through all spots using Algorithm~\ref{algo:search} with input arguments denoted in this paper as ``search interval'' $\Delta p$, ``initial location'' $X_0$, and ``preferred lane'' $Y_0$. The parameter $N_X$ is the total number of spots in $X$ direction when $Y$ is fixed.

\begin{algorithm}
\label{algo:search}
\caption{Spot Search}
    \SetArgSty{}
    \KwIn{$\Delta p$, $X_0$, $Y_0$, $N_X$}
    \KwOut{Spot Assignment $(x, y)$}
    
    \SetKwProg{Def}{def}{:}{}
    \Def{$\mathrm{SpotSearch}(\Delta p, X_0, Y_0)$}{
        $(x, y) = (X_0, Y_0)$\;
        \While{not all spots are assigned}{
            \uIf{$x \geq N_X$}{\label{algo:alc_back_1}
                \uIf{$\left[N_X \text{ mod } (\Delta p + 1) \right]$ = 0}{
                    $x \leftarrow (x + 1) \text{ mod } N_X$
                }
                \uElse{
                    $x \leftarrow x \text{ mod } N_X$ \label{algo:alc_back_2}
                }
            }
            \uIf{Spot $(x,y)$ is occupied}{
                $(x, y) \leftarrow (x, 1-y)$\;
            }
            \uElse{
                return $(x,y)$;
            }
            \uIf{Spot $(x,y)$ is occupied}{
                $(x, y) \leftarrow (x+1+\Delta p, Y_0) $\;
            }
            \uElse{
                return $(x,y)$;
            }
        }
    }
\end{algorithm}

Let the set ${\cal A}(k) = \left\{ {1,2,...} \right\}$ contains the vehicle indices at time $k$, ordered by arrival time. For vehicle $i \in {\cal A}(k)$ entering the parking lot, the spot $(x^i, y^i)$ is assigned according one of the three polices:
\begin{enumerate}[label=(\roman*)]
    \item \textit{``Random Search" (RS)}: A free spot is randomly picked without any preferences. 
    \item \textit{``Interval-first Search" (IS)}: Vehicles will prioritize a spot at least $\Delta p$ away from the front vehicle. $(x^i, y^i) = \mathrm{SpotSearch}(\Delta p, X_0=x^{i-1} + 1 + \Delta p, Y_0=Y^i_0)$.
    \item \textit{``Farthest-first Search" (FS)}: Vehicles will prioritize the farthest available spot from the gate. $(x^i, y^i) = \mathrm{SpotSearch}(\Delta p, X_0=0, Y_0=Y^i_0)$.
\end{enumerate}
where $Y^i_0$ is the $Y$ value of the spots next to its driving lane. When the search algorithm reaches $X = N_X$, it will loop back as described by line~\ref{algo:alc_back_1} - \ref{algo:alc_back_2} of Algorithm~\ref{algo:search}.

\begin{figure}[htbp]
\centering
\includegraphics[width=0.7\columnwidth]{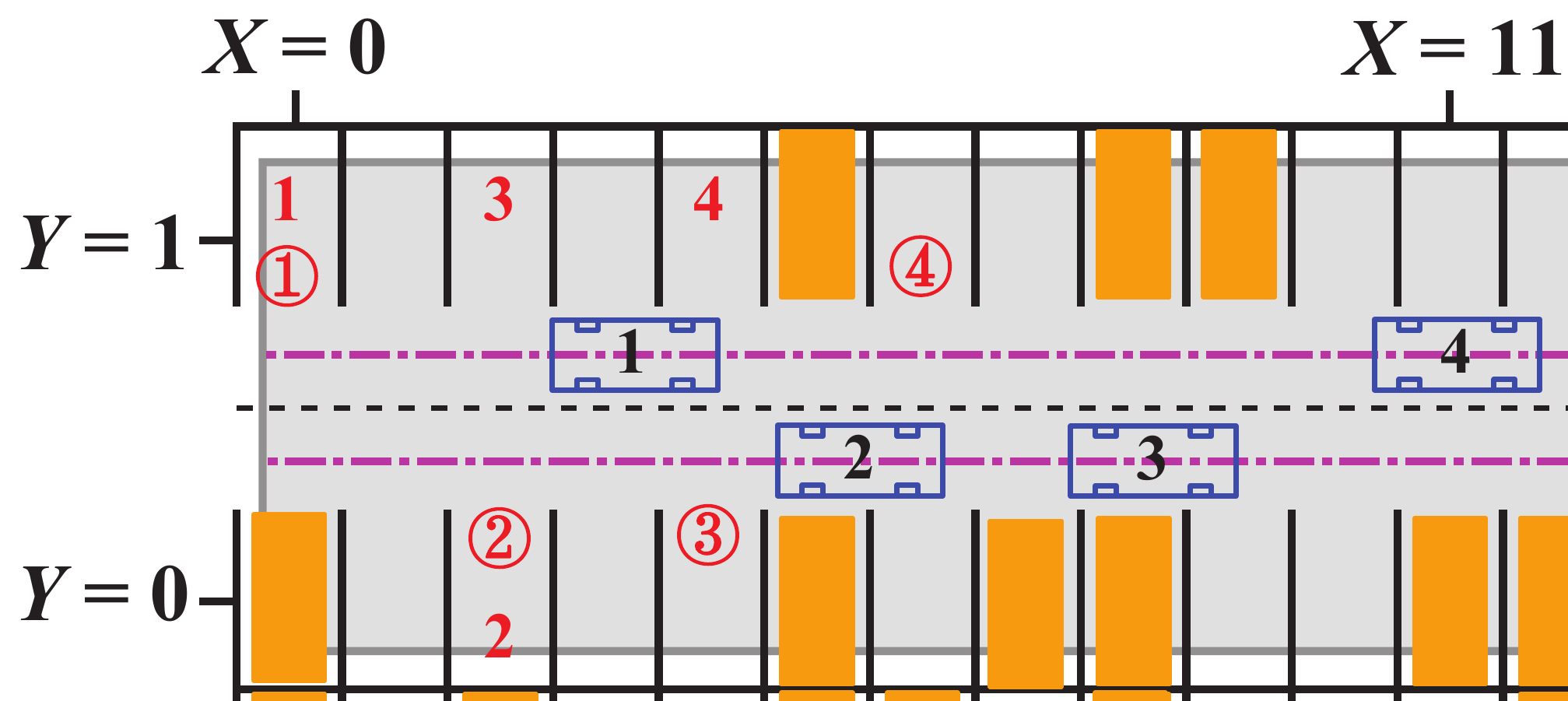}
\caption{Spots Assignment under IS and FS}
\label{fig:assign}
\vspace{-0.3cm}
\end{figure}

Fig.~\ref{fig:assign} takes the region of $X \in [0, 11]$ and 4 vehicles as an example. When $\Delta p = 1$, red circled numbers are spots assigned to the corresponding vehicles under IS policy, and red numbers without circles are assigned by FS policy.

\subsection{Path Generation}
\label{sec:path_generation}

A complete path for a vehicle is generated sequentially for two modes:
\subsubsection{Path for Queuing Mode}
The section of center line from the entrance to an end point near the target spot is selected as the operating trajectory.
\subsubsection{Path for Maneuvering Mode}
After reaching the end point of the queuing mode, a dynamically feasible trajectory starts from center line and leads the vehicle into the designated target without intruding into other spots.
These ``final leg'' maneuvering paths are generated by the Hierarchical Optimization-Based Collision Avoidance (H-OBCA) method~\cite{Zhang2017,Zhang2019} with a large set of possibles parameters, including the variation of starting positions and final poses. The resulting trajectories and input sequences are stored in an offline library for fast invoking when a certain maneuver is requested.

According to the assumption (A3) in Section~\ref{sec:setup}, a vehicle can be simplified to operate on the pre-planned path discussed above and modeled as the discrete-time linear model:
\begin{equation}
    s^{[i]}(k + 1) = s^{[i]}(k) + {v^{[i]}}(k){\Delta t},\forall i \in {\cal A}(k),
    \label{eq:dyn_model}
\end{equation}
where $\left[  {v^{[i]}}(k), {s^{[i]}}(k)\right]$ are longitudinal speed and position of the $i$-th vehicle and $\Delta t$ is the sampling time.


\subsection{Inter-vehicle Collision Avoidance}
\label{sec:safe_constr}
\begin{figure}[ht]
\centering
\includegraphics[width=0.9\columnwidth]{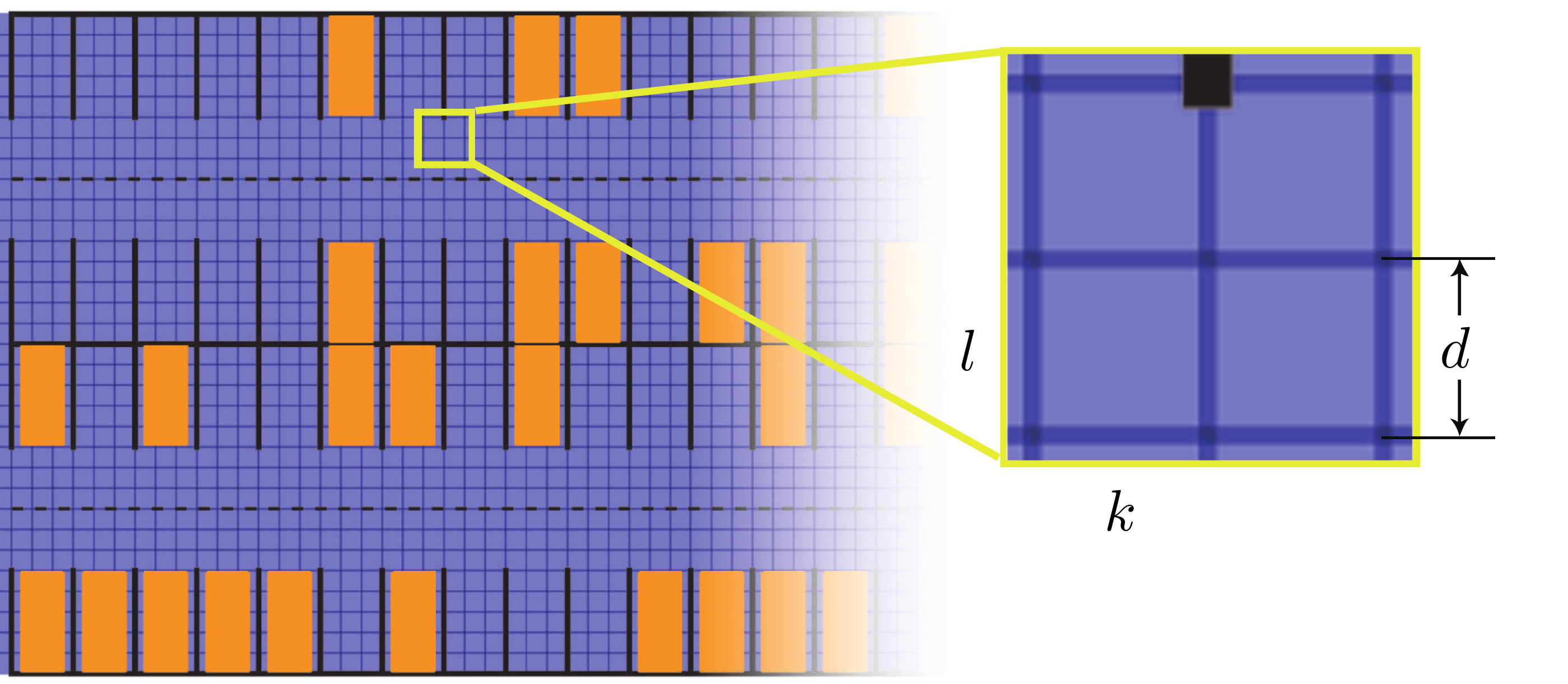}
\caption{Occupancy grid map. Orange spots contain parked vehicles.}
\label{fig:costmap}
\vspace{-0.3cm}
\end{figure}

The vehicles evaluate inter-vehicular interactions and avoid collisions using an occupancy grid map (Fig.~\ref{fig:costmap}) shared among all vehicles.
The occupancy grid is obtained by discretizing the parking lot uniformly with grid size $d$. As a result, the continuous space $\mathbb{R}^2$ is approximated by a grid space $\mathbb{Z}_+^2$. In practice, the level of discretization is often chosen in a trade-off between trajectory tracking tolerance requirement, perception sensor precision, and network bandwidth.

On the discrete grid, we denote by $\mathbb{B}(s^{[j]}(k)) \subset \mathbb{Z}_+^2$ the space occupied by the $j$-th vehicle at time $k$ during the queuing mode
(Fig.~\ref{fig:safety_constr_queue_1}). In maneuvering mode, we denote by $\mathbb{B}_M(s^{[l]}(k)) \subset \mathbb{Z}_+^2$ the space to be occupied by the $l$-th vehicle from time $k$ till the end of the maneuver (Fig.~\ref{fig:safety_constr_queue_2}).
The distinction between the two modes is necessary since vehicles require more space during their final parking maneuver and the trajectories are more complex.


The detailed safety constraints are discussed next.

\subsubsection{Safety in Queuing Mode}
\label{sec:safety_queue_mode}

\begin{figure}[htbp]
\centering
\subfigure[Constraint Eq.~(\ref{eq:queue_check_1})]{
\includegraphics[width=0.4\columnwidth]{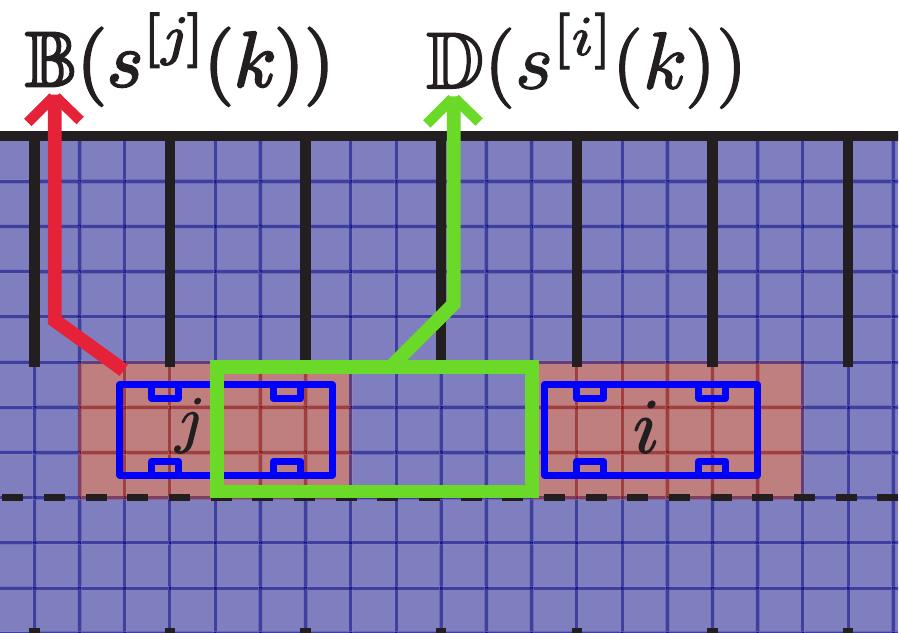}
\label{fig:safety_constr_queue_1}
}
\quad
\subfigure[Constraint Eq.~(\ref{eq:queue_check_2})]{
\includegraphics[width=0.4\columnwidth]{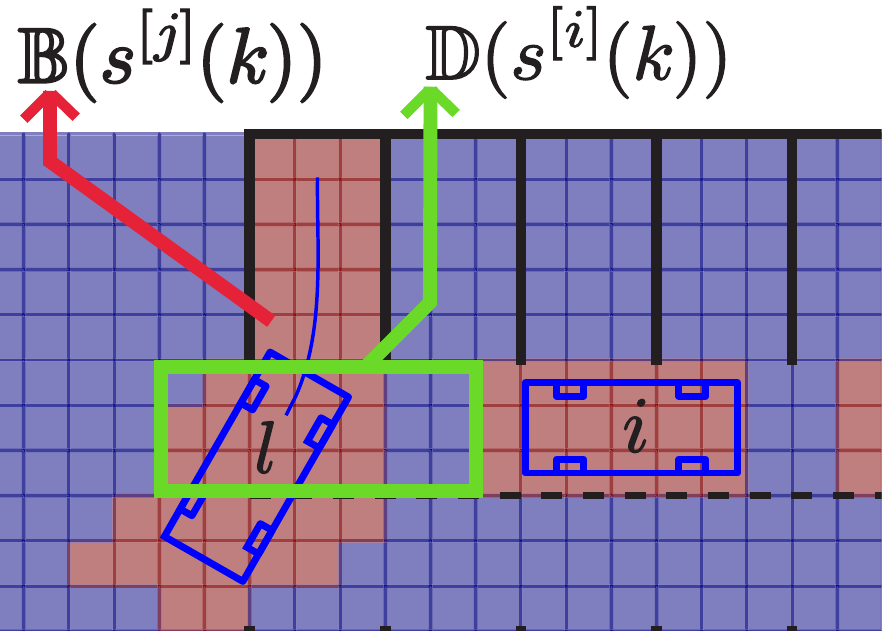}
\label{fig:safety_constr_queue_2}
}
\caption{Safety Constraints in Queuing Mode}
\label{fig:safety_constr_queue}
\end{figure}

When vehicle $i \in {\cal A}(k)$ is in queuing mode at time $k$, let $\mathbb{D}(s^{[i]}(k)) \subset \mathbb{Z}_+^2$ represent the forward reachable space, which contain all grids the vehicle occupies if it continues to move until time $k + \Delta K$, as contoured by green lines in Fig.~\ref{fig:safety_constr_queue}.
$\Delta K$ is a design parameter.

At each time $k$, the vehicle $i$'s decentralized control makes sure that $\mathbb{D}(s^{[i]}(k))$ does not intersect with the shape of any other vehicle $j$ in queuing mode. In addition, the vehicle $i$'s decentralized control ensures that $\mathbb{D}(s^{[i]}(k))$ does not intersect with the reachable space of vehicles $l$ that arrives earlier and already starts executing the final leg of a parking maneuver. Compactly the constraints can be written as:
\begin{subequations}
    \begin{align}
        \mathbb{D}(s^{[i]}(k)) \cap \mathbb{B}(s^{[j]}(k)) = \emptyset, & \forall j \in {\cal A}(k) \backslash i \label{eq:queue_check_1}, \\
        \mathbb{D} (s^{[i]}(k)) \cap \mathbb{B}_M(s^{[l]}(k)) = \emptyset, & \forall l \in {\cal A}(k) \backslash i, l < i \label{eq:queue_check_2},
    \end{align}
\end{subequations}
where vehicle $j, l$ are at $s^{[j]}(k),s^{[l]}(k)$. Examples of constraint violations are illustrated in Fig.~\ref{fig:safety_constr_queue} so vehicle $i$ must yield in these situations.

\subsubsection{Safety in Maneuvering Mode}
\begin{figure}[htbp]
\centering
\subfigure[Constraint Eq.~(\ref{eq:maneuver_check_1})]{
\includegraphics[width=0.4\columnwidth]{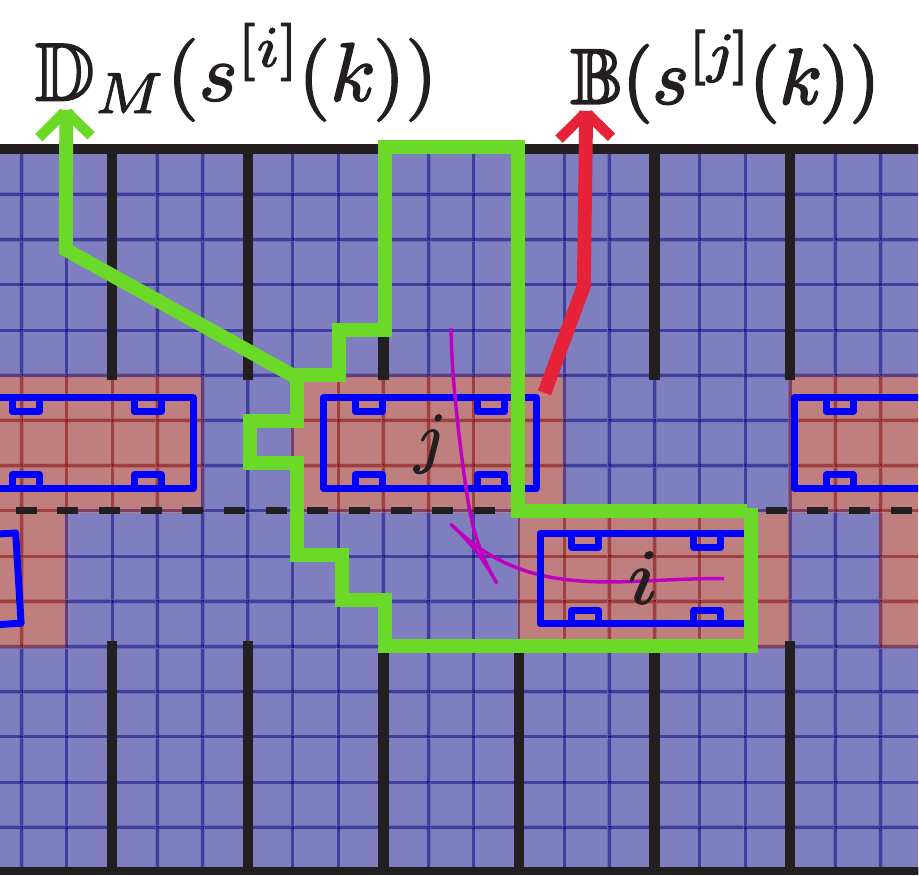}
\label{fig:safety_constr_manuever_1}
}
\quad
\subfigure[Constraint Eq.~(\ref{eq:maneuver_check_2})]{
\includegraphics[width=0.4\columnwidth]{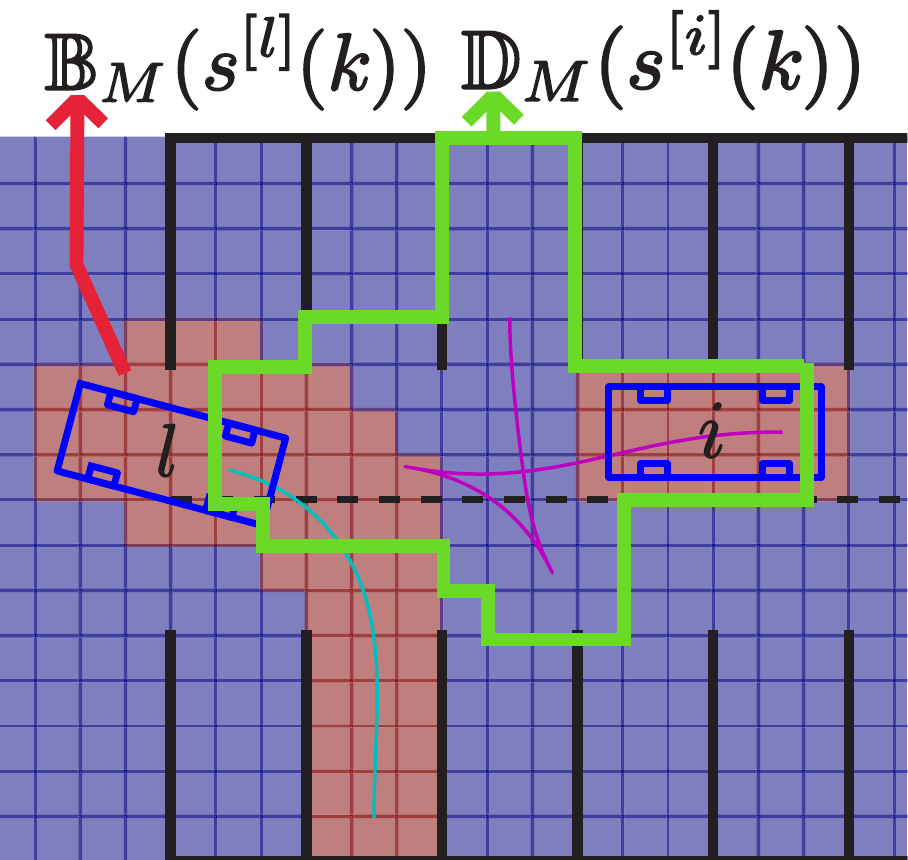}
\label{fig:safety_constr_manuever_2}
}
\caption{Safety Constraints in Maneuvering Mode}
\label{fig:safety_constr_manuever}
\end{figure}

When vehicle $i$ is in maneuvering mode at time $k$, let $\mathbb{D}_M(s^{[i]}(k)) \subset \mathbb{R}^2$ represent the forward reachable space from time $k$ to the end of the remaining maneuver, as contoured by green lines in Fig.~\ref{fig:safety_constr_manuever}. The safety constraints can be compactly written as:
\begin{subequations}
    \begin{align}
        \mathbb{D}_M(s^{[i]}(k)) \cap \mathbb{B}(s^{[j]}(k)) = \emptyset, & \forall j \in {\cal A}(k) \backslash i \label{eq:maneuver_check_1}, \\
        \mathbb{D}_M (s^{[i]}(k)) \cap \mathbb{B}_M(s^{[l]}(k)) = \emptyset, & \forall l \in {\cal A}(k) \backslash i, l < i \label{eq:maneuver_check_2}.
    \end{align}
\end{subequations}
Constraints (\ref{eq:maneuver_check_1}) and (\ref{eq:maneuver_check_2}) encode the fact that the  maneuver of vehicle $i$ cannot interfere with either the body of any other operating vehicle $j$ in queuing mode, or final leg maneuvers of other earlier-arrived vehicles $l$. The violation examples are shown in Fig.~\ref{fig:safety_constr_manuever} where vehicle $i$ must yield.

\begin{figure}[htbp]
\centering
\subfigure[Dead Lock]{
\includegraphics[width=0.32\columnwidth]{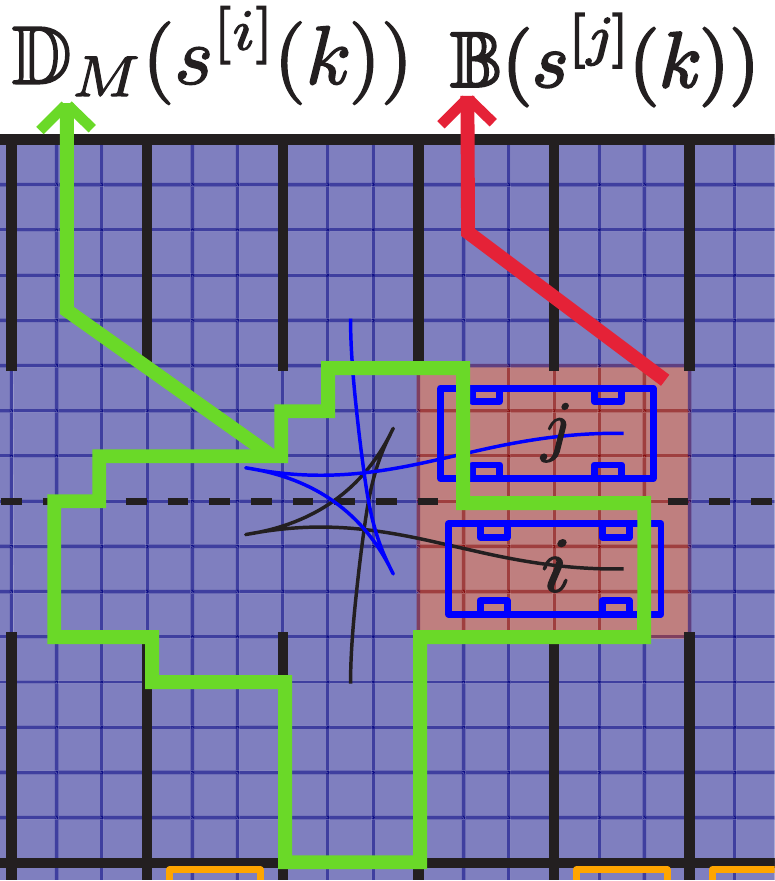}
\label{fig:deadlock}
}
\quad
\subfigure[Resolved]{
\includegraphics[width=0.32\columnwidth]{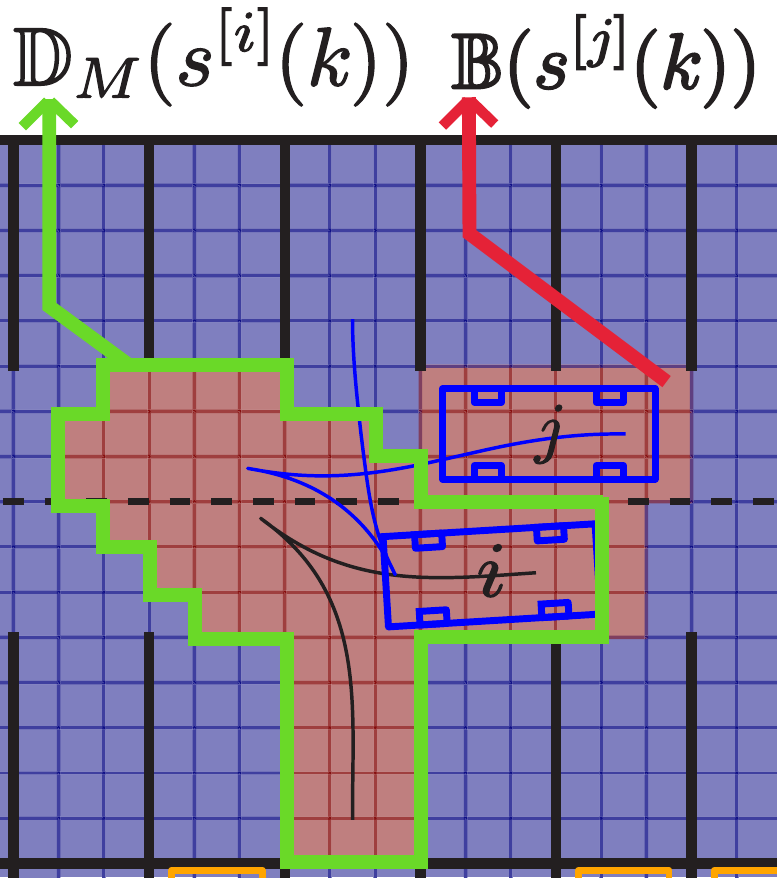}
\label{fig:deadlock_resolved}
}
\caption{Deadlock and Resolution}
\label{fig:deadlock_N_resolution}
\vspace{-0.5cm}
\end{figure}

Enforcing constraints Eq.~(\ref{eq:maneuver_check_1})-(\ref{eq:maneuver_check_2}) at all times will lead to infeasible deadlock scenarios as the one depicted in Fig.~\ref{fig:deadlock}
, where Eq.~(\ref{eq:maneuver_check_1}) is violated for both vehicles at the same time:
\begin{subequations}
    \begin{align}
        & \mathbb{D}_M (s^{[i]}(k)) \cap \mathbb{B}(s^{[j]}(k)) \neq \emptyset, \\
        & \mathbb{D}_M (s^{[j]}(k)) \cap \mathbb{B}(s^{[i]}(k)) \neq \emptyset.
    \end{align}
\label{eq:deadlock}
\end{subequations}
By generating a new deadlock-free maneuver for vehicle that arrives earlier, the situation can be resolved as Fig.~\ref{fig:deadlock_resolved}. 
\vspace{-0.3cm}


\subsection{Complete Control Algorithm}
The complete control algorithm is described in Algorithm~\ref{algo:control}. Every time a new vehicle arrives, it will be allocated to a driving lane, assigned a target spot, and generated a corresponding path to track. The algorithm is capable of not only coordinating vehicles safely in both queuing and maneuvering modes, but also resolving infeasible deadlocks occurred. The speed $v^{[i]}(k)$ of each vehicle $i \in {\cal A}(k)$ at time $k$ will be sent to the low-level controller.

Note that although this work mainly deals with the behavior of incoming vehicles, departing vehicles can be analyzed by ``reversing" the trajectories in maneuvering mode and exiting the parking lot in queuing mode. The constraints can be kept. 

\begin{algorithm}
\label{algo:control}
\caption{Control Algorithm}
\SetArgSty{}
    \KwIn{New Vehicle Arrivals}
    \KwOut{$v^{[i]}(k), \forall i \in {\cal A}(k)$ at each time $k$}
    
    Initialize occupancy grids\;
    Initialize time $k = 0$\;
    \Repeat{${\cal A}(k) = \emptyset$}{
        Reset occupancy grids\;
        Add new-arriving vehicles into ${\cal A}(k)$, allocate as Section~\ref{sec:allocate_strategy}, and plan path as Section~\ref{sec:path_generation}\;
        \For{$\forall i \in {\cal A}(k)$}{
            Occupy the vehicle body $\mathbb{B}(s^{[i]}(k))$\;
        }
        \For{$\forall i \in {\cal A}(k)$}{
            \uIf{In maneuvering mode}{
                Check the safety constraints Eq.~(\ref{eq:maneuver_check_1})-(\ref{eq:maneuver_check_2})\;
                \uIf{Collision free}{
                    Occupy the ongoing maneuver $\mathbb{B}_M(s^{[i]}(k))$\;
                }
            }
        }
        \For{$\forall i \in {\cal A}(k)$}{
            \uIf{In queuing mode}{
                Check the safety constraint Eq.~(\ref{eq:queue_check_1})-(\ref{eq:queue_check_2})\;
            }
            \uElseIf{In maneuvering mode}{
                Check the safety constraints Eq.~(\ref{eq:maneuver_check_1})-(\ref{eq:maneuver_check_2})\;
            }
            \uIf{Collision free}{
                Proceed by outputting $v^{[i]}(k) = v_{\mathrm{ref}}$\;
            }
            \uElse{
                Yield by outputting $v^{[i]}(k) = 0$\;
            }
        }
        \uIf{Deadlock happens}{
            Resolve by regenerating a feasible maneuver\;
        }
        \For{$\forall i \in {\cal A}(k)$}{
            \uIf{Vehicle $i$ has finished the parking task}{
                ${\cal A}(k+1) \leftarrow {\cal A}(k) \backslash i$\;
            }
        }
        $k \leftarrow k+1$\;
    }
\end{algorithm}

\section{Simulation Result and Discussion}
\label{sec:simulation}
\subsection{Simulation Parameters}
In this section, we illustrate Algorithm~\ref{algo:control} on the parking lot shown in Fig.~\ref{fig:Overview}. 
The parking lot has a length of $66\si{m}$ and a width of $16\si{m}$, and 
each spot is of size $5\si{m} \times 3\si{m}$. 
We assume that 48 spots out of 88 are available, whose locations are randomly chosen.
The vehicles are modeled as rectangles of size $4.7\si{m} \times 2 \si{m}$.
The grid size is chosen to be $1\si{m} \times 1\si{m}$ to fit the parking lines and vehicle dimensions. 
When moving forward, the reference speed is $v_{\mathrm{ref}} = 4 \si{m/s}$. The sampling time is $\Delta t = 0.1 \si{s}$. 
The preview horizon, used to compute the forward reachable space presented in Section~\ref{sec:safety_queue_mode}, is chosen as $\Delta K = 15$ steps. Furthermore, we make the following assumptions:
\begin{enumerate}[label=(\roman*)]
\item Following the literature~\cite{Ross2010}, the arrival times of the vehicles are exponentially distributed with mean $1/\lambda$;

\item If both lanes are open (denoted as ``2L"), then the vehicles randomly chose the lane they drive on. Vehicles will not have choices if only one lane (``1L") is open.

\item The final parking maneuvers are randomly chosen to be either forward parking or reverse parking.
\end{enumerate}

In the following, we consider arrival times with parameters $1/\lambda \in \{ 1,2,4,7\} \si{s}$. For the policies IS and FS, the investigated search intervals are $\Delta p \in \{ 0,1,...,21\}$. This results in a total of 360 possible combinations (4 arrival rates; 2 lane opening choices; 3 assignment policies, where FS and IS each has 22 $\Delta p$ values). We executed 100 simulations runs for settings that use the IS or FS policy, and 2200 simulations runs with RS policy to marginalize the randomness associated to the random spot assignment. The demonstration of some chosen scenarios can be found at {\tt\small http://bit.ly/fleetpark}.


\subsection{Simulation Results}
We first focus on the scenario where only one lane (1L) is open, and present results with two lanes later on. To evaluate the effectiveness of the algorithms, we introduce two metrics: Mean Task Time (MTT), and Maximum Queue Length (MQL), which we describe next.

\subsubsection{Mean Task Time (MTT)}
\begin{figure}[htbp]
\centering
\subfigure[$1/\lambda = 1$]{
\includegraphics[width=0.44\columnwidth]{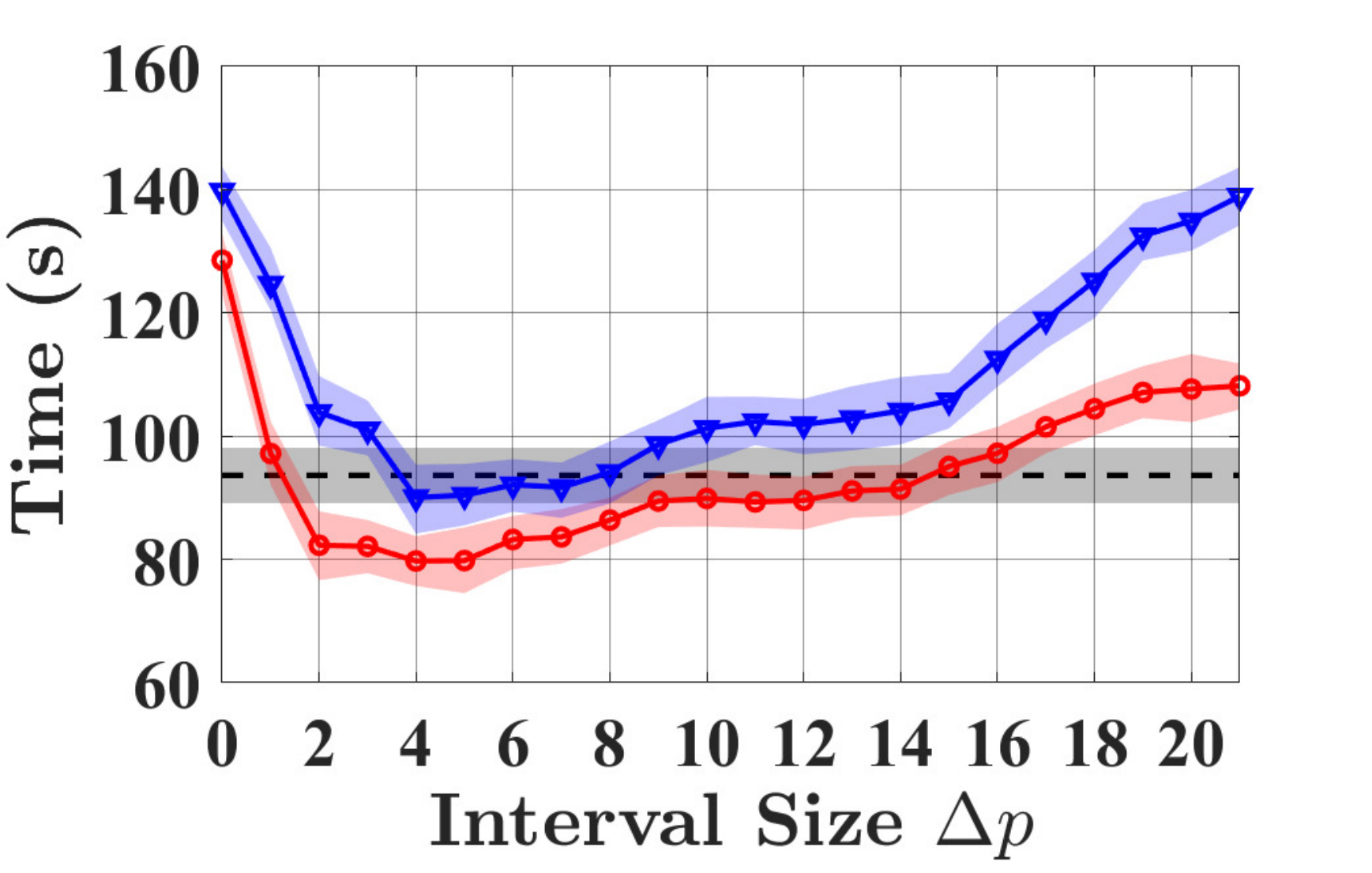}
}
\subfigure[$1/\lambda = 2$]{
\includegraphics[width=0.44\columnwidth]{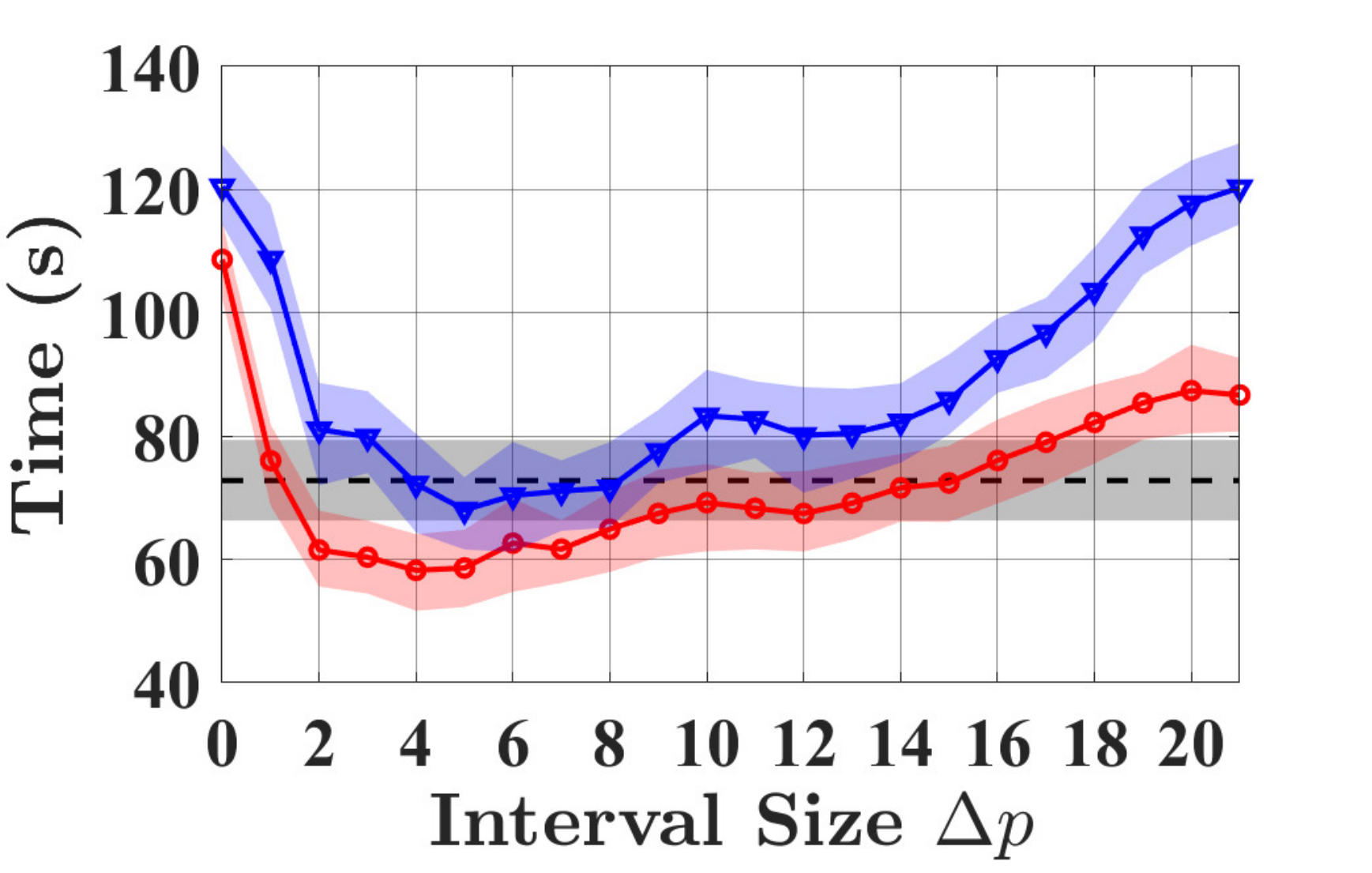}
}
\subfigure[$1/\lambda = 4$]{
\includegraphics[width=0.44\columnwidth]{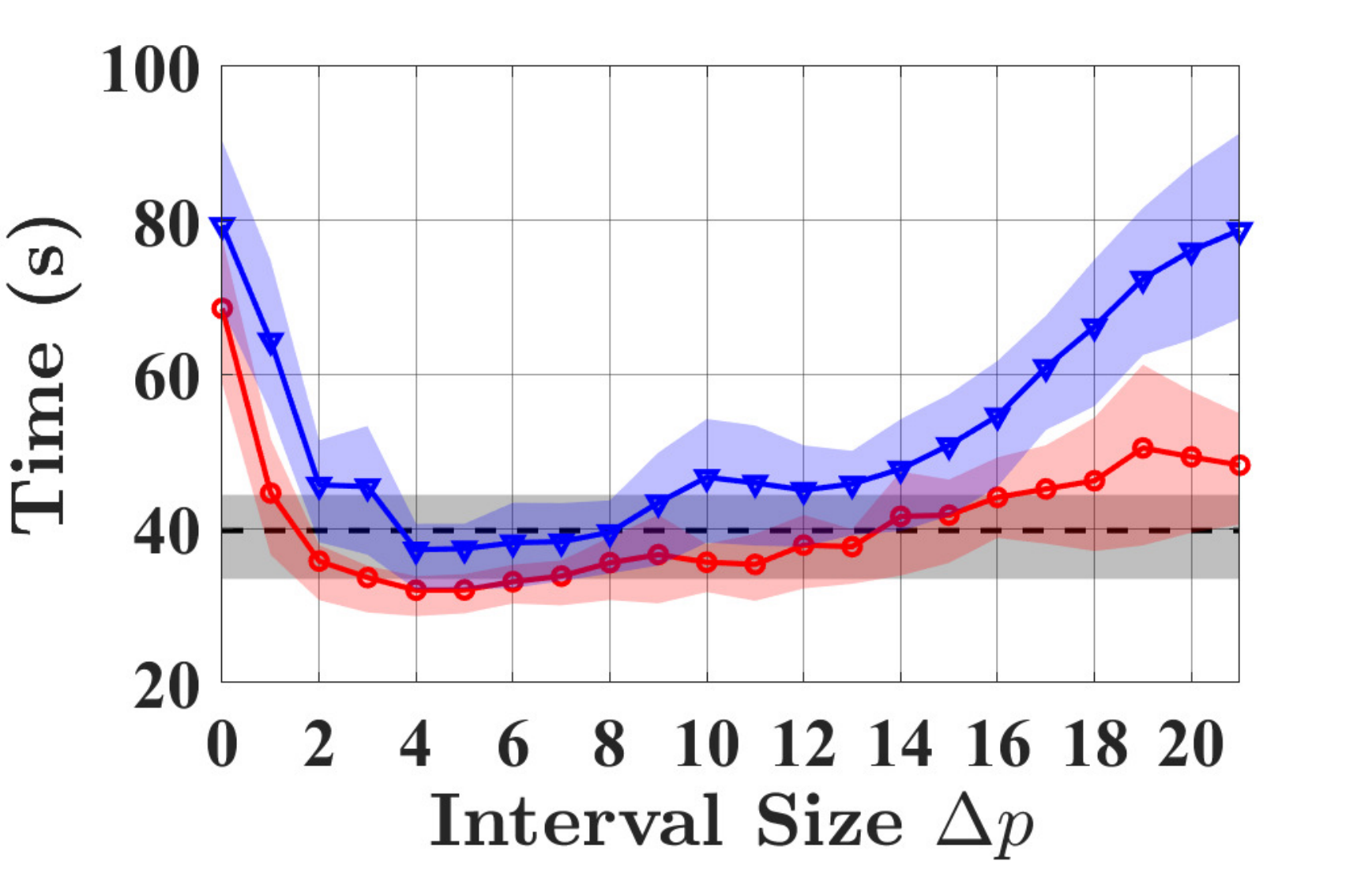}
}
\subfigure[$1/\lambda = 7$]{
\includegraphics[width=0.44\columnwidth]{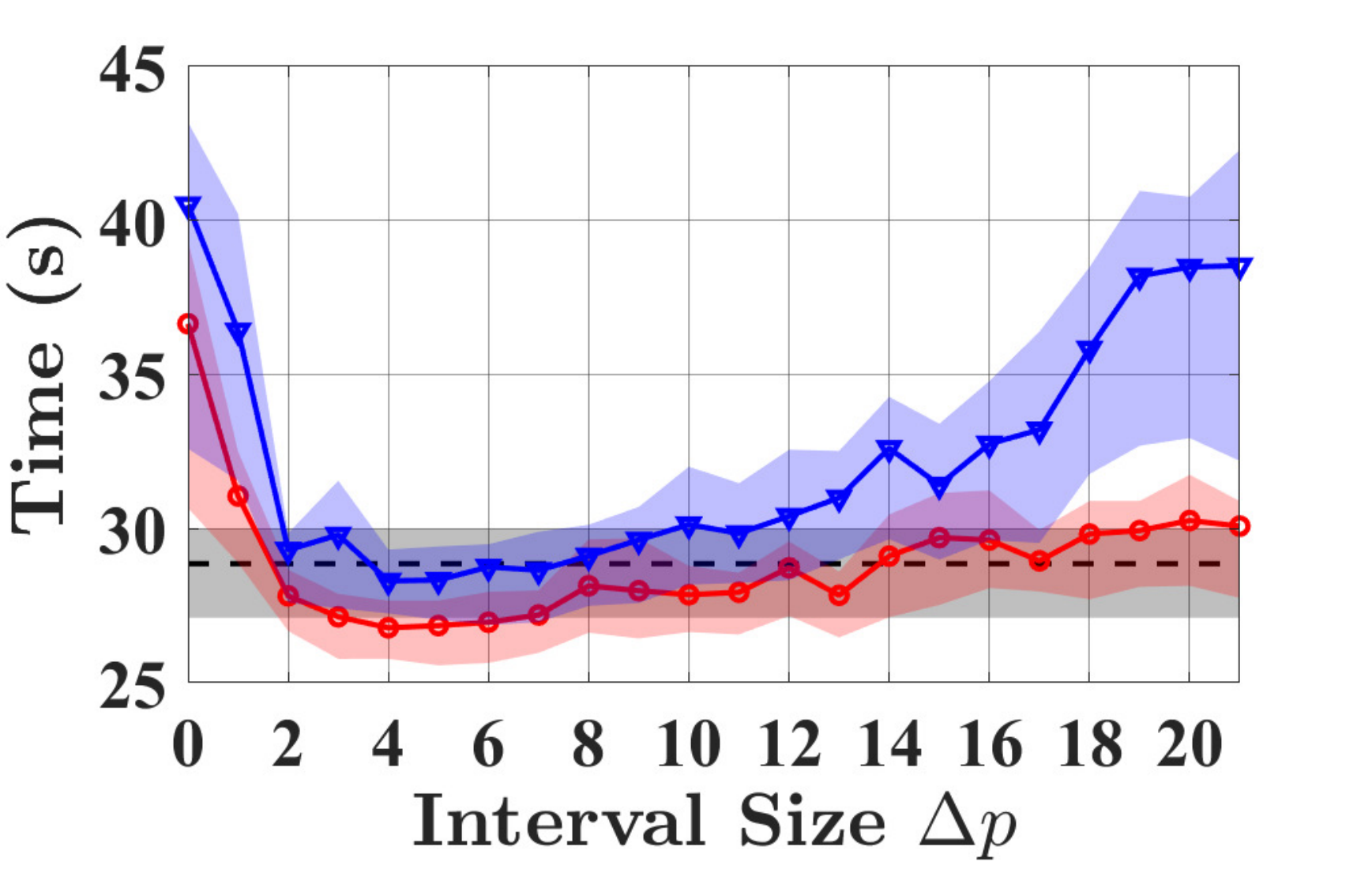}
}
\caption{$t_\textnormal{MTT}\left( \lambda, \Delta p \right)$ with one lane open (1L). Black dashes (-), red circles (\textcolor{red}{$\circ$}), and blue triangles (\textcolor{blue}{$\triangledown$}) denote RS, \textcolor{red}{IS}, and \textcolor{blue}{FS} policies respectively. Subfigures use {different} scaling.}
\label{fig:MTT_1L}
\end{figure}

The Mean Task Time $t_\textnormal{MTT}\left( \lambda, \Delta p \right)$ is defined as the average time length a vehicle spends to finish parking with respect to the specified arrival rate and search interval. Formally, 
\begin{equation}
t_\textnormal{MTT}\left( \lambda, \Delta p \right) = \frac{1}{N}\sum_{i = 1}^{N} \left( t^{[i]}_{\textnormal{f}, \lambda, \Delta p} - t^{[i]}_{\textnormal{0}, \lambda, \Delta p} \right),
\end{equation}
where $t^{[i]}_{\textnormal{0}, \lambda, \Delta p}, t^{[i]}_{\textnormal{f}, \lambda, \Delta p}$ are the time that $i$ enters the parking lot and completes parking, and $N$ is the total number of vehicles. For the operator, it is desirable to keep MTT low to finish the parking task as soon as possible.
The results are shown in Fig.~\ref{fig:MTT_1L}, where lines are averages over all simulation runs and shaded regions are the interquartile range.

We make the following observations:
\begin{enumerate}[label=(\roman*)]
    \item The MTT is higher for higher arrival rates (i.e., when $1/\lambda$ is smaller). This is intuitive because the parking lot is more congested when vehicles arrive at a faster rate, reducing the free space the vehicles can maneuver in.
    \item For all arrival rates, the FS policy always exhibits higher MTT values than the IS policy, and even higher than the random strategy RS. 
    This is because, FS policy does not guarantee an appropriate interval between two consecutive vehicles (see vehicle $2,3$ in Fig.~\ref{fig:assign}), and the resulting conflicts will increase the parking time.
    \item For all arrival rates, both IS and FS policies have the highest MTT values when $\Delta p = 0$ or $21$. This is intuitive because, in such scenarios, vehicles will spend a lot of time either trying to avoid collisions,
    or simply have to queue outside of the parking lot. Only a few of them can execute the maneuver successfully at a time.
    \item For all arrival rates, both IS and FS policies achieve the lowest MTT when $\Delta p \approx 4$. 
    This is because vehicles need space while maneuvering, and we see from Fig.~\ref{fig:Overview} that the space required approximately corresponds to two spots.
    Therefore, $\Delta p \approx 4$  provides sufficient space for adjacent vehicles to maneuver simultaneously.
\end{enumerate}

\subsubsection{Maximum Queue Length (MQL)}
\begin{figure}[htbp]
\centering
\subfigure[$1/\lambda = 1$]{
\includegraphics[width=0.44\columnwidth]{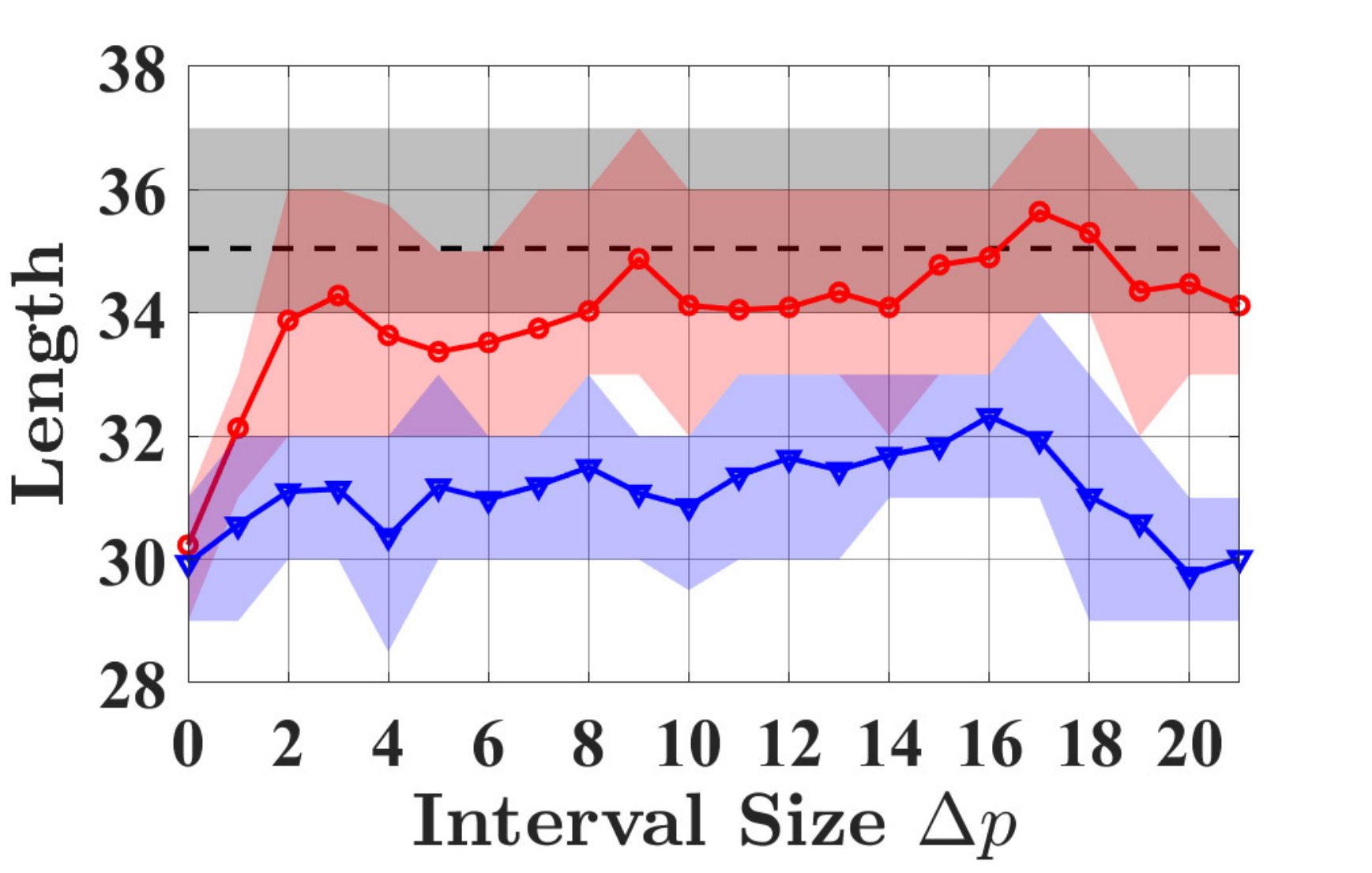}
\label{fig:MQL_1L_1}
}
\subfigure[$1/\lambda = 2$]{
\includegraphics[width=0.44\columnwidth]{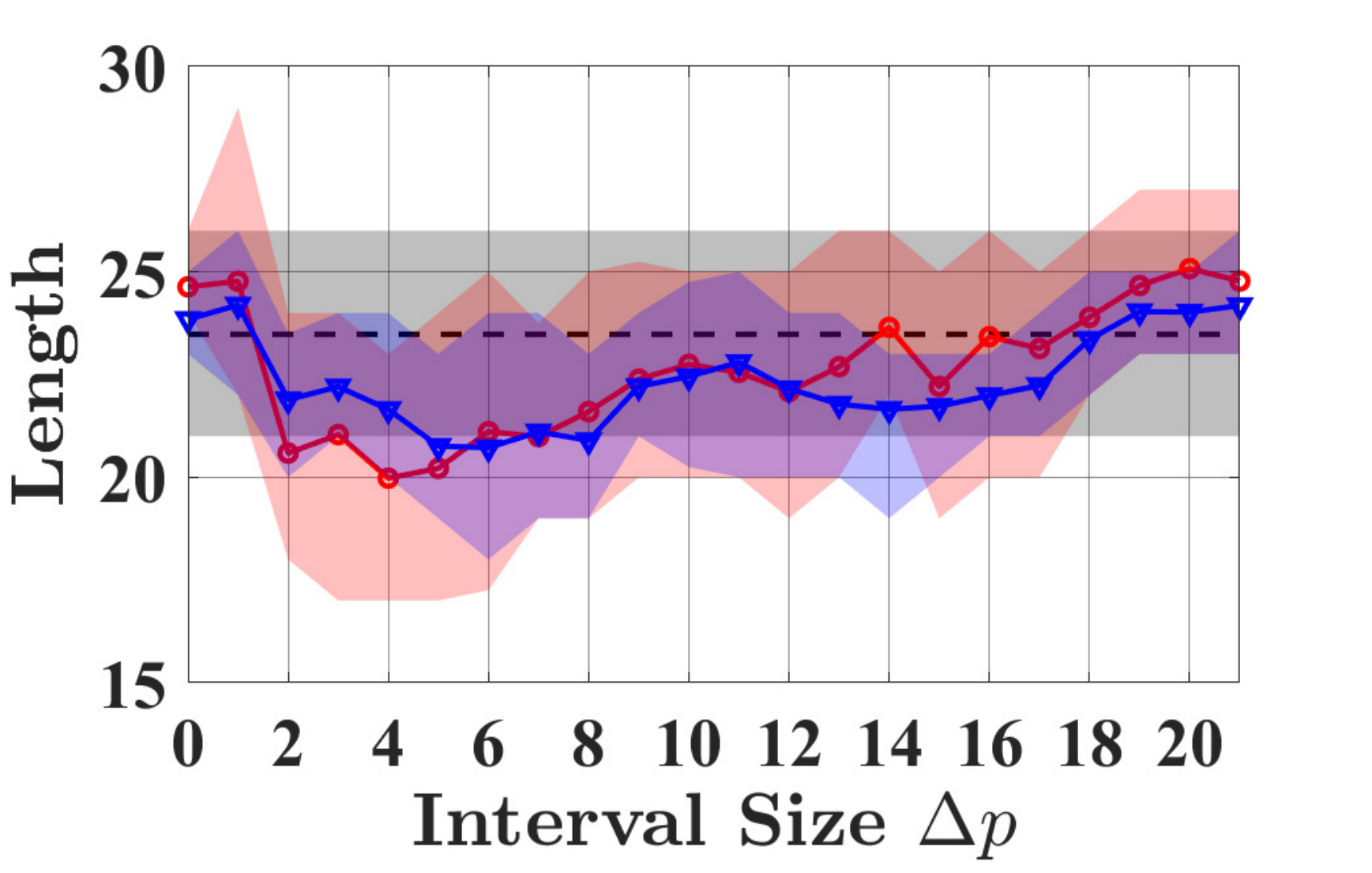}
\label{fig:MQL_1L_2}
}
\subfigure[$1/\lambda = 4$]{
\includegraphics[width=0.44\columnwidth]{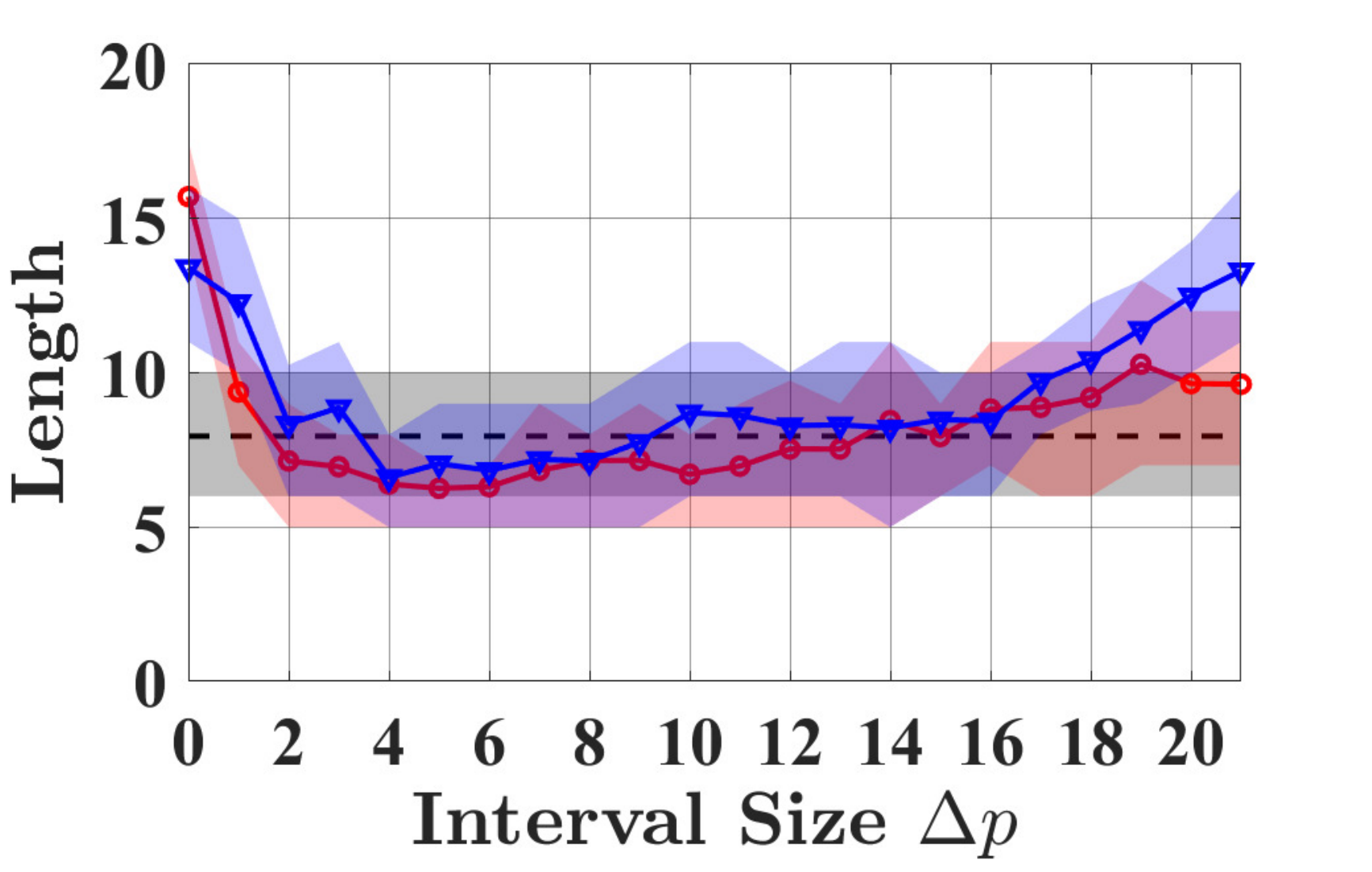}
\label{fig:MQL_1L_4}
}
\subfigure[$1/\lambda = 7$]{
\includegraphics[width=0.44\columnwidth]{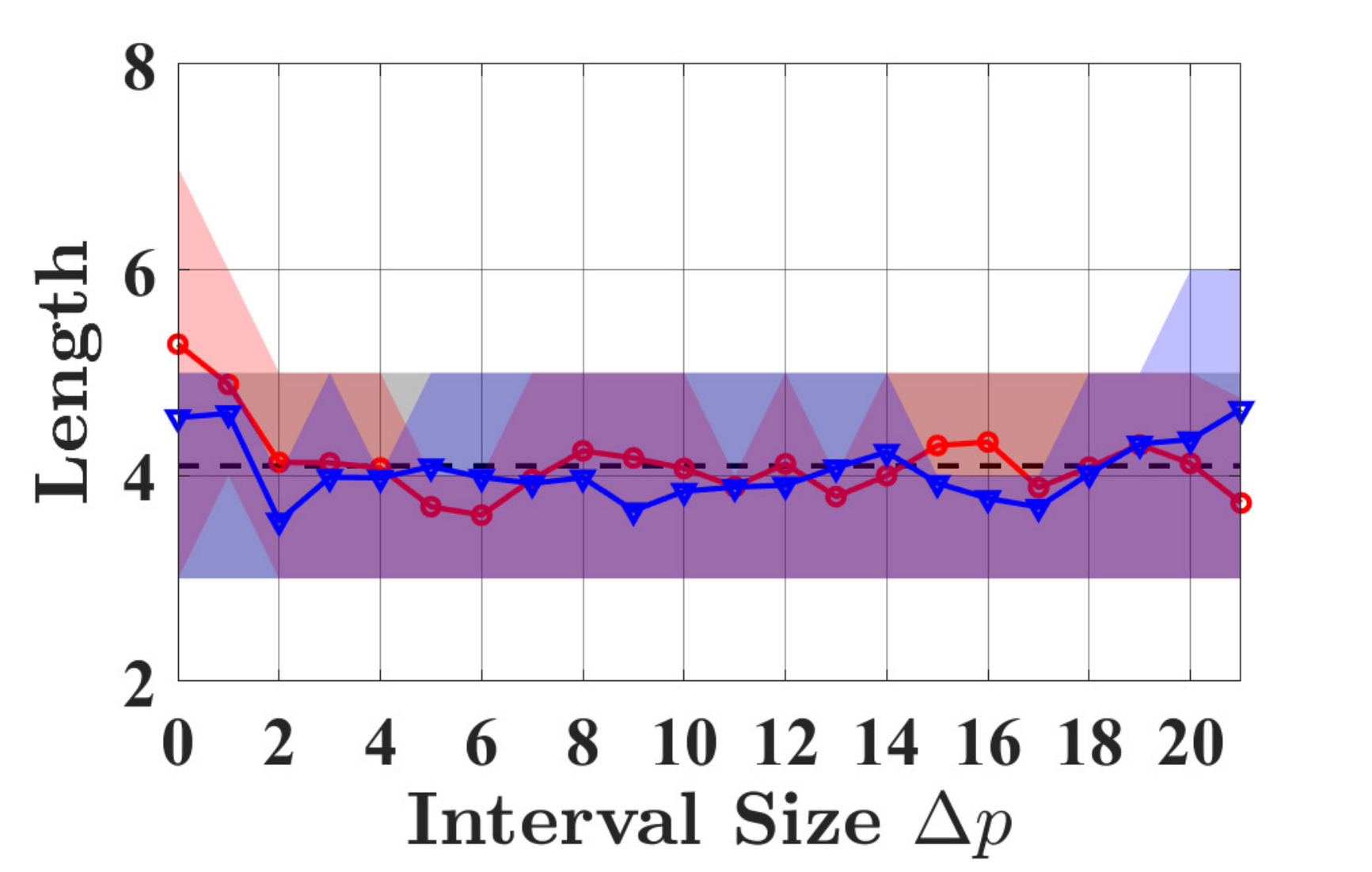}
\label{fig:MQL_1L_7}
}
\caption{$l_\textnormal{MQL}\left( \lambda, \Delta p \right)$ with one lane open (1L). Black dashes (-), red circles (\textcolor{red}{$\circ$}), and blue triangles (\textcolor{blue}{$\triangledown$}) denote RS, \textcolor{red}{IS}, and \textcolor{blue}{FS} policies respectively. Subfigures use different scaling.}
\label{fig:MQL_1L}
\end{figure}

We define the Maximum Queue Length $t_\textnormal{MTT}\left( \lambda, \Delta p \right)$ as the
maximum number of vehicles waiting outside the parking lot with respect to the specified arrival rate and search interval, from the first arriving vehicle to the last one completing its parking task. Formally,
\begin{equation}
l_\textnormal{MQL}\left( \lambda, \Delta p \right) = \max_{k \geq 0} \sum_{i\in {\cal A}(k)} I(s^{[i]}_{\lambda, \Delta p}(k))
\end{equation}
where $I(s^{[i]}_{\lambda, \Delta p}(k))=1$ if the the $i$-th vehicle is outside the parking lot, and $0$ else.
From a ``social" perspective, it is desirable to keep MQL low as not to disturb surrounding traffic.
The simulation results are shown in Fig.~\ref{fig:MQL_1L} with average value and interquartile range.

We make the following observations:
\begin{enumerate}[label=(\roman*)]
    \item The higher the arrival rate, the higher is MQL, since more vehicles then will wait to enter the parking lot.
    \item When the  arrival rate is high (e.g., $1/\lambda=1$), the FS policy achieves lower MQL values than IS policy. 
    This is because the ``farthest-first" characteristic of FS will push more vehicles into the farthest end of parking lot and reduce the queue length outside. As expected, this effect becomes subtler when vehicles arrive at a slower rate, i.e., when $1/\lambda$ is bigger.
    \item In contrast to MTT, it is difficult to identify a unique parameter for $\Delta p$ that achieves the lowest MQL for all values of $\lambda$. However, averaged over the tested $1/\lambda$, it seems that $\Delta p=4$ is a reasonable choice, achieving the best ``overall" value.
\end{enumerate}

\begin{figure}[htbp]
\centering
\subfigure[$t^*_\mathrm{MTT}(\lambda)$]{
\includegraphics[width=0.44\columnwidth]{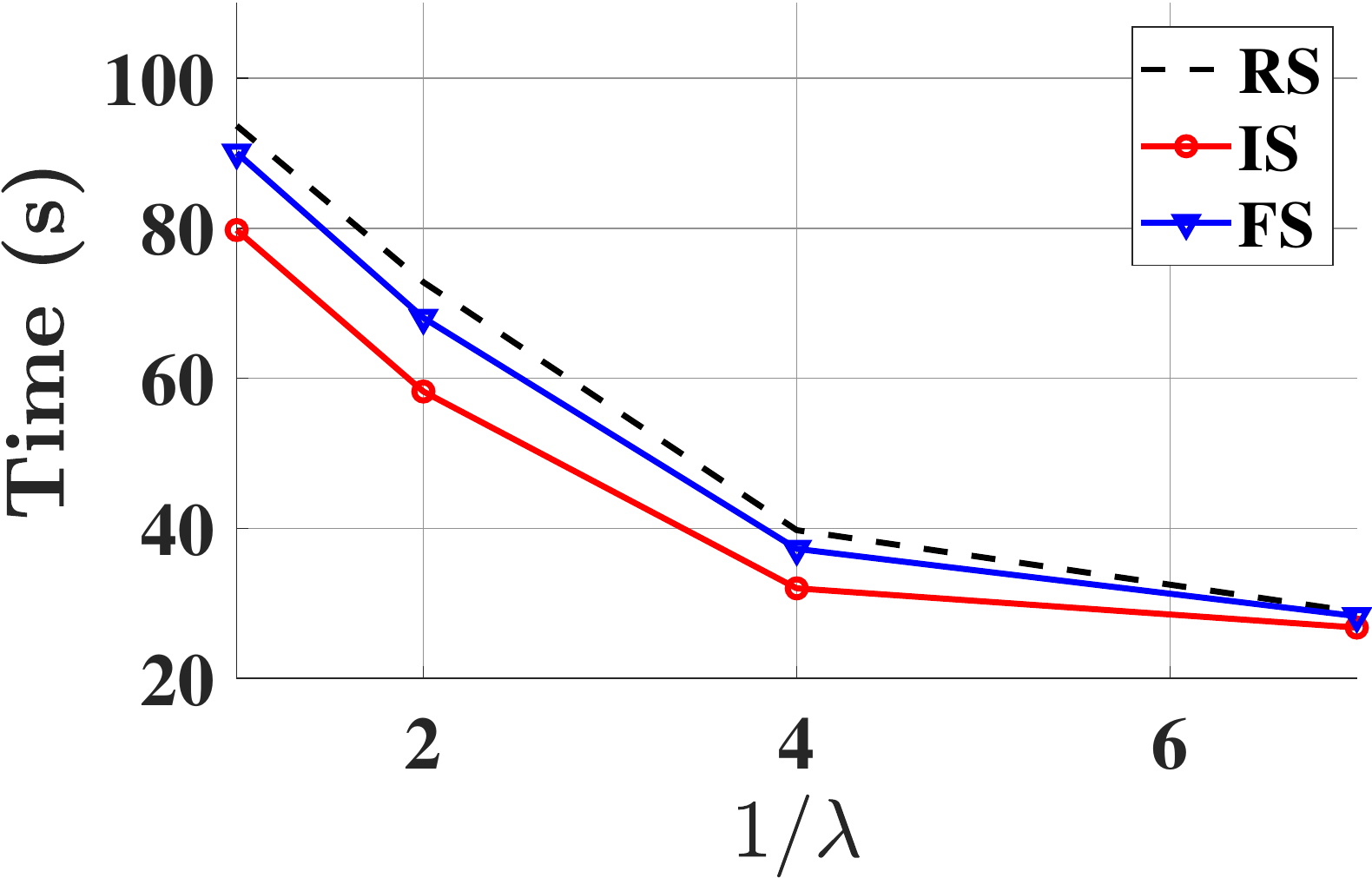}
}
\subfigure[$l^*_\mathrm{MQL}(\lambda)$]{
\includegraphics[width=0.44\columnwidth]{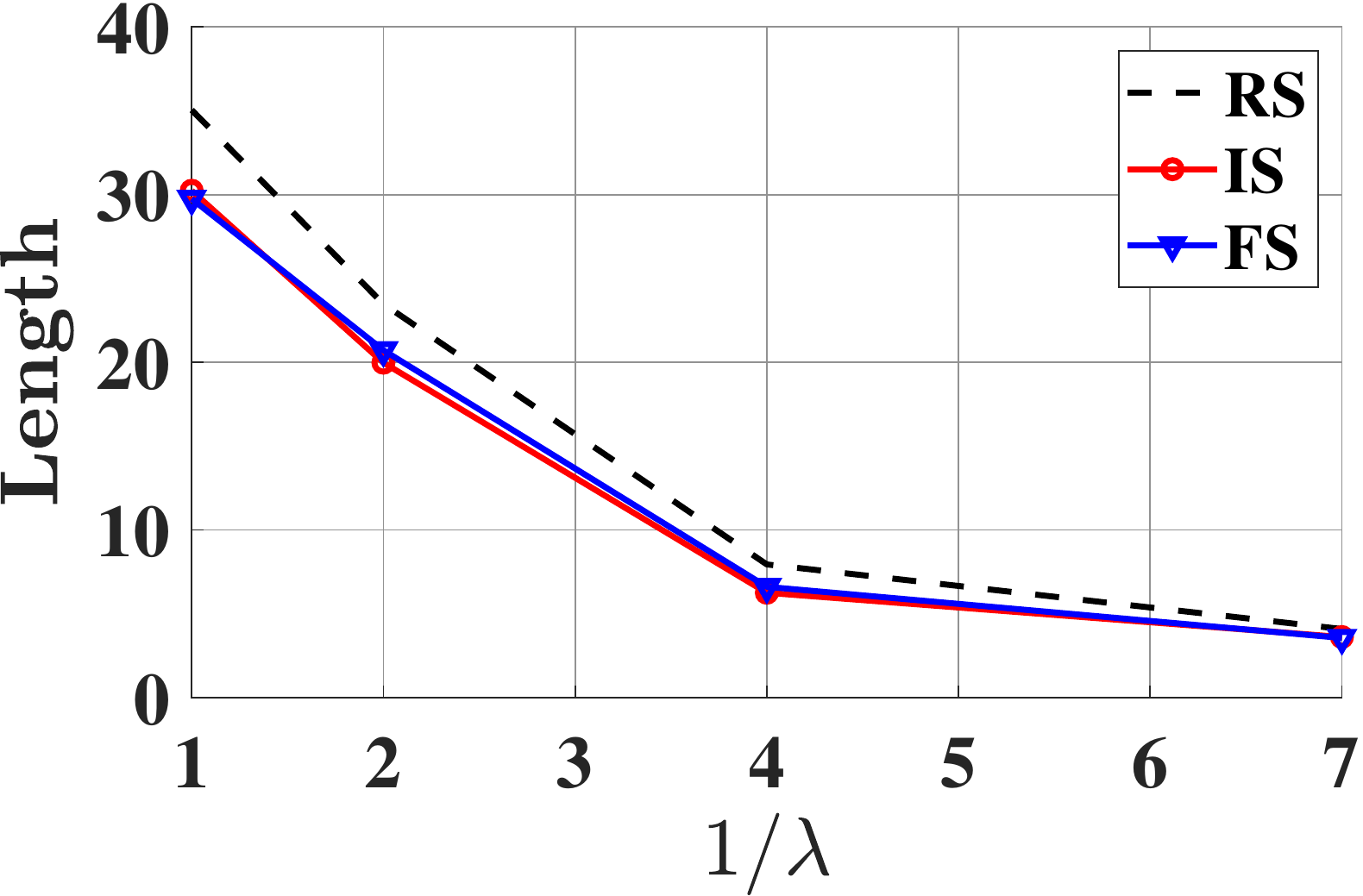}
}
\caption{Optimal Average Value with 1L}
\label{fig:Opt_1L}
\vspace{-0.3cm}
\end{figure}

Fig.~\ref{fig:Opt_1L} depicts the lowest MTT and MQL under different arrival rates for the case when only one lane is open (1L):
\begin{subequations}
\begin{align}
    & t^*_\mathrm{MTT}(\lambda) = \min_{\Delta p} t_\mathrm{MTT}(\lambda, \Delta p), \\
    & l^*_\mathrm{MQL}(\lambda) = \min_{\Delta p} l_\mathrm{MQL}(\lambda, \Delta p).
\end{align}
\end{subequations}
It indicates that the IS performs better than the FS and the RS policy. Specifically, in terms of MTT, IS achieves improves upon RS up to 20\%, hand up to 14\% when compared to the FS policy. 
Both IS and FS policies behave similarly in terms of MQL, outperforming RS by up to 21\%.

\subsection{Two-Lane (2L) Scenario}
\begin{figure}[htbp]
\centering
\subfigure[$t^*_\mathrm{MTT}(\lambda)$]{
\includegraphics[width=0.44\columnwidth]{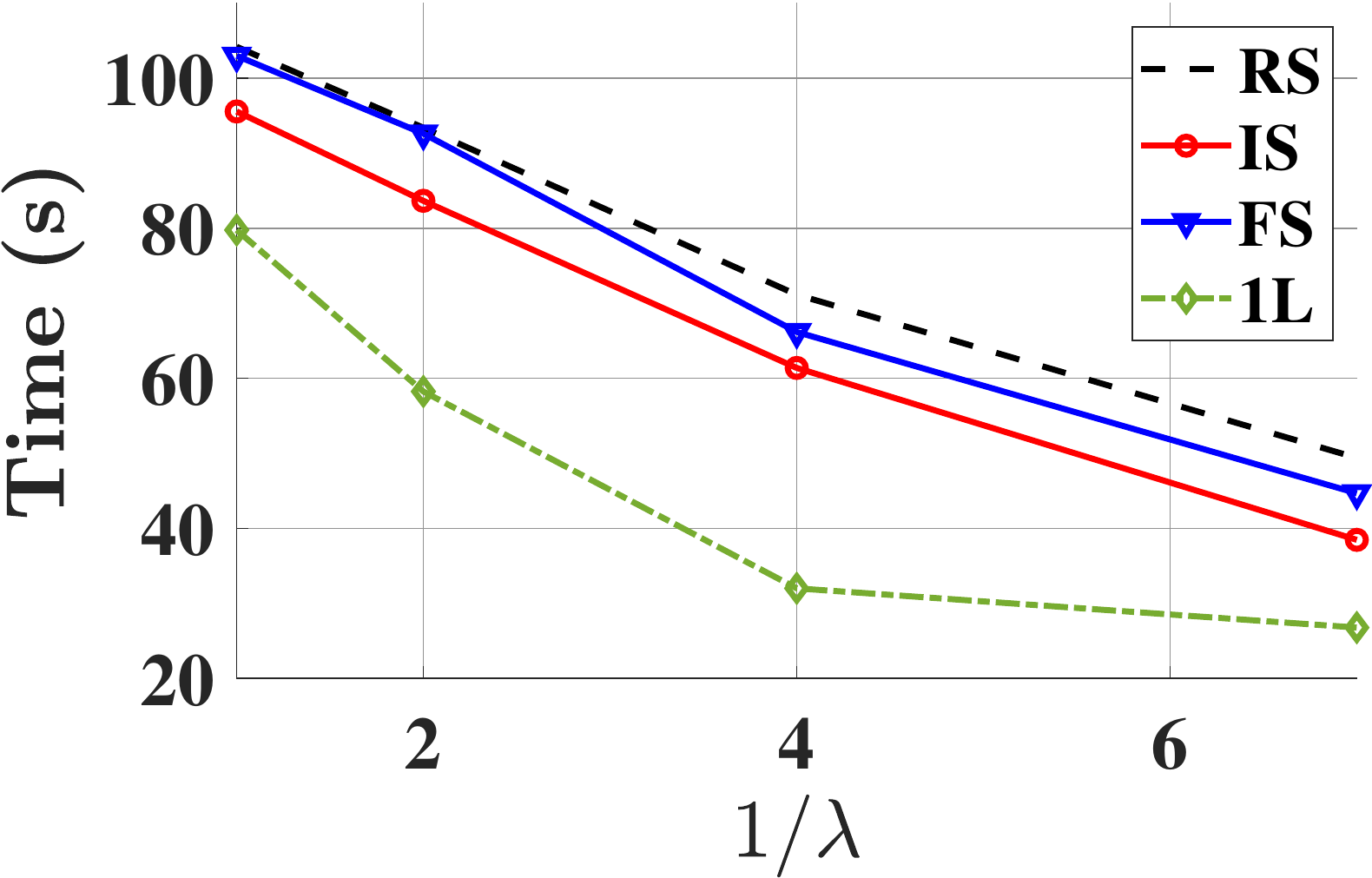}
}
\subfigure[$l^*_\mathrm{MQL}(\lambda)$]{
\includegraphics[width=0.44\columnwidth]{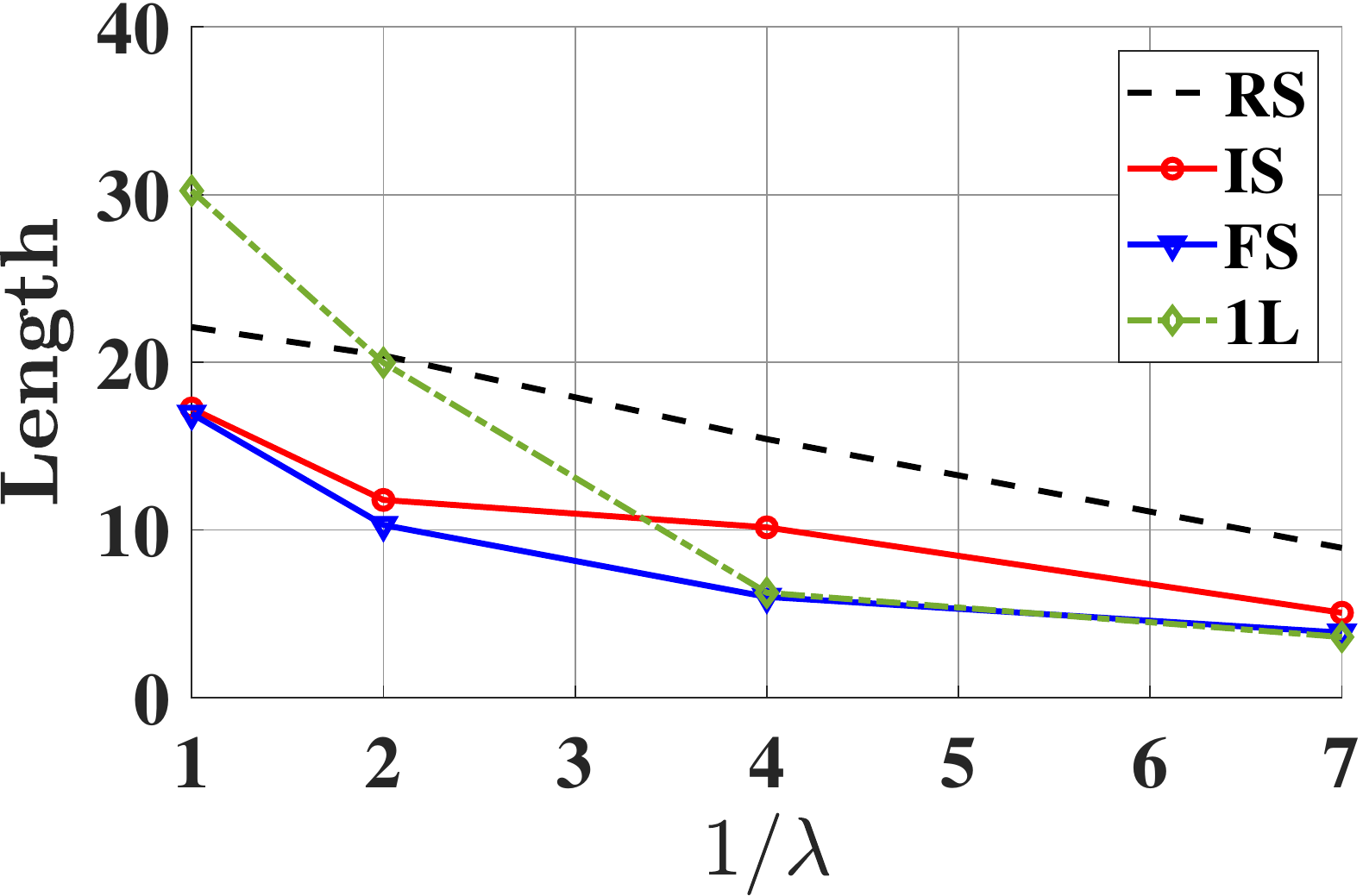}
\label{fig:Opt_2L_MQL}
}
\caption{Optimal Average Value with 2L. Green dash-dot line with diamonds (\textcolor{mygreen}{$\diamondsuit$}) denotes the best possible performance  with 1L (from Fig.~\ref{fig:Opt_1L}).}
\label{fig:Opt_2L}
\vspace{-0.3cm}
\end{figure}

In this section, we present results for the case when both lanes are open (2L) and Fig.~\ref{fig:Opt_2L} shows $t^*_\mathrm{MTT}(\lambda)$ and $l^*_\mathrm{MQL}(\lambda)$.
Compared to the 1L case (green line),
we see that all three policies perform worse in terms of MTT, i.e., vehicles do not park faster when both lanes are open.
Our conjecture is that, due to the tight geometry of the parking lot, vehicles need to intrude into both lanes when executing their maneuvers and others may yield more often to avoid collision. Therefore, opening two lanes slows down the parking process, similar to what is known as Braess's paradox~\cite{MURCHLAND1970391} in road networks. 
In term of queue length, we see from Fig.~\ref{fig:Opt_2L_MQL} that FS policy outperforms both the IS and RS policy, improving upon RS by up to 53\%, and results in shorter queues compared to 1L. This is intuitive, as 2L has more space than 1L and FS searches for the farthest spots, hence can hold more vehicles inside the parking lot lanes.

\section{Conclusion}
\label{sec:conclude}
In this work, we studied the problem of autonomous parking of a large fleet of vehicles inside a parking lot.
We proposed a hierarchical system-level framework that is able to handle large numbers of vehicles in a computationally efficient way. Our algorithm solves the spot allocation and path generation centrally, while collision avoidance is tackled by the vehicles individually in a decentralized fashion, enabling scalability.

Extensive numerical simulations confirm our intuition that different parking lot allocation strategies have significant impact on the fleet parking time and the queue length. For example, when the objective is to minimize the fleet parking time, then opening one lane (1L) and selecting spots that are far enough for vehicles to park simultaneously
(IS policy) leads to the best performance. On the other hand, if the objective is to minimize queue length, then opening two lanes (2L) and choosing parking spots that are in the farthest end of parking lot
(FS policy) gives the best solution, but comes at the cost of higher parking time.
Finally, our simulation results reveal that Braess's paradox applies to parking lots as well; this observation should be taken into account when designing parking lots and parking algorithms for large vehicle fleets.

\renewcommand{\baselinestretch}{0.9}

\addtolength{\textheight}{-12cm}

\bibliographystyle{ieeetr}
\bibliography{References.bib}

\end{document}